\documentclass[10pt,conference]{IEEEtran}
\IEEEoverridecommandlockouts
\usepackage[T1]{fontenc}
\usepackage{newtxtext}
\usepackage{newtxmath}
\usepackage{graphicx}
\usepackage{booktabs}
\usepackage{array}
\usepackage{xurl}
\usepackage{amsmath}
\usepackage{pifont}
\usepackage{tikz}
\usetikzlibrary{arrows.meta,positioning}
\newcommand{\cmark}{\ding{51}}
\newcommand{\xmark}{--}
\newcolumntype{L}[1]{>{\raggedright\arraybackslash}p{#1}}

\newif\ifarxiv
\arxivtrue

\begin{document}

\title{Broken on Arrival: Non-Functional Quantized LLM Artifacts in Public Model Registries and How to Catch Them%
\ifarxiv\thanks{This work has been submitted to the IEEE for possible publication. Copyright may be
transferred without notice, after which this version may no longer be accessible.}\fi}

\author{\IEEEauthorblockN{Aditi Patodiya}
\IEEEauthorblockA{Independent Researcher\\
Email: aditi.patodia31@gmail.com}}

\maketitle

\begin{abstract}
Developers download quantized large language models (LLMs) from public registries to run them locally.
None of the registries we examined documented functional testing of their files. We tested all 305
eligible code-capable GGUF tags (the llama.cpp format) across 15 model lines in the official Ollama
library, plus 22 popular Hugging Face community files. A 15-task code screen identified suspects, which we retested on all 164
HumanEval+ tasks. Seven official tags (2.3\%) solved at most six of them. All were low-bit quantizations
of models with at most four billion parameters. We rebuilt five suspect files from the original weights.
All five registry files matched our uncalibrated rebuilds tensor for tensor, so none was corrupted. Exact
reproducibility did not imply function. With the weights, toolchain and tensor mix held fixed, adding an
importance matrix raised Qwen2.5-Coder-3B scores at three quantization levels from zero to 76--113 of 164 tasks. The same
uncalibrated recipe left the smaller Qwen2.5-Coder-1.5B functional, showing that the recipe alone does
not predict failure. None of 66 sampled tags that passed the screen was non-functional. We release the
audit data and the quantcheck tool and recommend screening low-bit tags before publication, then
calibrating or withholding failures.
\end{abstract}


\section{Introduction}
Running a large language model (LLM) locally now usually means pulling a quantized file from a public
registry. An internet-wide measurement found more than 320{,}000 publicly reachable self-hosted LLM
services, including 155{,}423 instances of Ollama, a widely used local LLM
runtime~\cite{hou2025deployment}. A year-long measurement saw 152{,}137 distinct IP addresses exposing
Ollama~\cite{xu2026ollamawild}. The downloaded file is a conversion of the model its authors evaluated, typically
stored in GGUF at 2 to 8 bits per weight. GGUF is the single-file model format of llama.cpp, the
open-source inference engine on which Ollama is built~\cite{llamacpp}. Registry operators or community
volunteers produce these conversions, and neither the registry pages nor the file metadata we examined
record a functional test before users download them.

The gap is not hypothetical. In 2024 llama.cpp maintainers found that byte-pair-encoding (BPE) GGUF
files converted before their fix fall back to a default pre-tokenization that is wrong ``in almost all
cases''~\cite{gerganov2024bpe}, and users of a curated registry then reported affected
models~\cite{sealad2024ollama}. In our census two years later, 26 of that registry's 305 eligible tags
still lack a pre-tokenizer type. Quantization research also shows that aggregate metrics hide damage to
individual answers~\cite{dutta2024accuracy,marchisio2024multilingual}. What has not been measured across a
registry is how often its quantized files fail to work, and why.

Viewed as a data pipeline, model distribution moves a checkpoint through conversion, quantization and
publication without a recorded functional check. For downstream users, a registry tag is therefore an
unverified input. This study supplies the checks such a pipeline lacks. It verifies the
published files by executing them at registry scale, establishes provenance by tracing failing files to
their source checkpoint and recipe at the tensor level, and tests reproducibility by rebuilding them
exactly. Execution and rebuilding answer different questions, because a file can match its recipe
exactly and still not work.

We execute 327 code-capable artifacts, including every eligible tag of 15 model lines in the official
Ollama library. A 15-task screen, tuned on known-healthy builds, identifies suspects, which we rerun on
all 164 HumanEval+ tasks~\cite{liu2023evalplus}. We use \emph{recipe-matched rebuilds} to identify why the
five Qwen and phi3.5 suspects fail. For each, we quantize the original checkpoint ourselves with the
recipe the file's metadata indicates, then compare the distributed file with our rebuild tensor for
tensor and task by task.

The answer surprised us. Seven official artifacts are non-functional, yet none of the five we rebuilt is
a corrupted file. They are exact, uncalibrated quantizations of the released weights. For
Qwen2.5-Coder-3B, adding an importance matrix raises HumanEval+ scores from 0\% to 46--69\% at three
of its four failing levels. The source weights, toolchain and tensor mix are held fixed in this
comparison. Section~\ref{sec:bg} explains the importance matrix. The first version of this study~\cite{patodiya2026v1} chose its reference conversions by
distributor rather than by recipe and reported these tags as defective conversions. The recipe-matched
rebuild corrects that diagnosis.

Four research questions (RQs) structure the study:
\begin{itemize}
\item \textbf{RQ1 (prevalence):} how many distributed artifacts are non-functional, and where?
\item \textbf{RQ2 (cause):} is a failure due to the file, the recipe, the model, the chat template, or the backend?
\item \textbf{RQ3 (detection without executing generated code):} can cheap output statistics or perplexity flag the failures?
\item \textbf{RQ4 (acceptance gate):} does the 15-task screen work as a registry acceptance test?
\end{itemize}

The contributions are:
\begin{itemize}
\item \textbf{An execution census with a released audit dataset.} We test 305 eligible tags of an official registry, 291 of them quantized, plus 22 community files, and release every census transcript.
\item \textbf{A recipe-matched attribution procedure.} Demonstrated on five failing artifacts, it compares a failing file with a same-recipe rebuild from source and separates file defects from recipe, model and backend effects.
\item \textbf{Evidence that exact reproducibility does not imply function.} Five failing registry files match our uncalibrated rebuilds tensor for tensor. For Qwen2.5-Coder-3B, one calibration step lifts three of the four failing levels to 46--69\% of HumanEval+, against 80\% for its 4-bit tag.
\item \textbf{A screen with measured limits.} Screening takes about 30\,s of generation per artifact. None of the 66 sampled tags that passed the screen is non-functional, and the five rebuilt failures remain flagged under our suite-size and generation-settings checks. We release the screen with the attribution scripts as the quantcheck tool.
\end{itemize}

\section{Background and Related Work}\label{sec:bg}
\subsection{Quantization and calibration}
GGUF ``k-quants'' quantize weights in
blocks with per-block scales and assign each tensor a type by a fixed
mix~\cite{kawrakow2023kquants,sparrenberg2025llamacpp}. A level such as \texttt{q3\_K\_M} names this mix
rather than a single bit width. Levels are lowercase for registry tags and uppercase
(\texttt{Q3\_K\_M}) for llama.cpp builds, following each tool's spelling. Calibration methods use sample activations to guide weight quantization.
Examples include GPTQ~\cite{frantar2023gptq} and activation-aware weight quantization
(AWQ)~\cite{lin2024awq} for checkpoints, and the llama.cpp importance matrix (imatrix) for
k-quants~\cite{kawrakow2024imatrix}. The imatrix holds activation statistics from a calibration text,
which llama.cpp uses to weight the rounding error when it chooses block scales. Its author reports the
largest perplexity gains at the lowest bit widths~\cite{kawrakow2024imatrix}, and Qwen's quantization
guide recommends AWQ scales or an imatrix and notes that sometimes a mixture ``just does not
work''~\cite{qwen2025guide}. For code models, 4-bit AWQ preserves correctness and code
quality~\cite{afrin2025dealbreaker}. Other studies compare the accuracy cost of post-training
quantization methods~\cite{giagnorio2025quant,afrin2026confidence} and find that code-specific calibration
data limits the loss at 3 and 2 bits~\cite{giagnorio2025quant}. Small models trained on many tokens
degrade most under low-bit quantization~\cite{ouyang2025undertrained}.

Isolated reports of the failure we study exist. A llama.cpp discussion reports gibberish from an
uncalibrated \texttt{Q2\_K} build of Qwen2.5-3B~\cite{llamacpp12378}. SynConfRoute served
Qwen2.5-Coder-3B at \texttt{q2\_K} through Ollama, measured 0\% on a fill-in-the-middle benchmark, and
attributed it to the precision level~\cite{thangarajah2026synconfroute}. Closest to our census, Singh et
al.\ probed the quantization ladders of 13 Ollama registry models and 14 Unsloth builds on Hugging Face
(HF). They
found the registry's \texttt{q2\_K} builds of Phi-3, Phi-3.5-mini (the tag we flag) and Qwen2.5-3B
unusable, and left open whether those files were mis-quantized or the models not compressible to
that level~\cite{singh2026saying}. None of these studies compares plain and calibrated
builds of the same weights at the same level, or rebuilds a failing registry file to recover its recipe.

\subsection{Execution audits and provenance}
Studies of model hubs mine metadata, inspect files statically and sometimes execute models to examine
reuse practices~\cite{jiang2023reuse}, naming conventions~\cite{jiang2025naming},
maintenance~\cite{castano2024evolution}, and security risks~\cite{jiang2022artifacts,zhao2024poisoning}.
Wang et al.\ set out an LLM supply-chain research agenda~\cite{wang2024agenda}. Model equality testing compares an API's output
distribution with a reference model's to detect changes such as quantization~\cite{gao2025equality}. A study of substitution audits finds output-based checks unreliable
against quantized or partial substitution~\cite{cai2025substitution}. Kimi Vendor Verifier checks
third-party deployments of one model family~\cite{moonshot2026kvv}. Kadasi et al.\ re-evaluate 500 hub
models against their model cards~\cite{kadasi2025hubs}, and Yao et al.\ execute the example code in 291
model cards~\cite{yao2026looksgood}.

Twenty-eight Ollama-library tags, down to 3 bits,
have been benchmarked for accuracy, latency and energy on a Raspberry Pi~\cite{husom2025sustainable},
and Kurt compares llama.cpp formats after in-house conversion~\cite{kurt2026quantization}. Quantization can also be
made to trigger malicious behavior~\cite{egashira2024exploiting}, including with GGUF
k-quants~\cite{egashira2025gap}. Closest to our attribution step, Liu and Tsai
test whether HF GGUF files can be re-derived bit for bit from the publishers' own full-precision GGUF
files~\cite{liu2026verifiability}. Their question is
whether a file can be reproduced. Ours is whether it works and, if not, why. Table~\ref{tab:related}
summarizes the comparison.

\begin{table}[t]
\centering
\caption{Execution studies. \emph{Census}: every eligible tag of an official registry.
\emph{Recipe}: controls for the quantization recipe. \emph{Cause}: controlled attribution of functional
failure to file, recipe, model or backend.}
\label{tab:related}
\setlength{\tabcolsep}{2.2pt}
\footnotesize
\begin{tabular}{@{}llccc@{}}
\toprule
Work & Object executed & Census & Recipe & Cause \\
\midrule
Gao et al.~\cite{gao2025equality} & API endpoints & \xmark & \xmark & \xmark \\
Cai et al.~\cite{cai2025substitution} & simulated substitution & \xmark & \xmark & \xmark \\
Kimi Vendor Verifier~\cite{moonshot2026kvv} & vendor deployments & \xmark & \xmark & \xmark \\
Kadasi et al.~\cite{kadasi2025hubs} & hub task models & \xmark & \xmark & \xmark \\
Husom et al.~\cite{husom2025sustainable} & Ollama tags (edge) & \xmark & \xmark & \xmark \\
Kurt~\cite{kurt2026quantization} & own conversions & \xmark & \cmark & \xmark \\
SynConfRoute~\cite{thangarajah2026synconfroute} & Ollama tags & \xmark & \xmark & \xmark \\
Singh et al.~\cite{singh2026saying} & Ollama and HF builds & \xmark & \xmark & \xmark \\
Liu and Tsai~\cite{liu2026verifiability} & HF GGUF rebuilds & \xmark & \cmark & \xmark \\
This work & registry tags, rebuilds & \cmark & \cmark & \cmark \\
\bottomrule
\end{tabular}
\end{table}

\subsection{Acceptance testing in software supply chains}
Package registries met a related problem first. Supply-chain attacks have been
classified~\cite{ladisa2023sok}, weak links in npm measured~\cite{zahan2022npm}, and malicious packages
shown to reach users through normal installs~\cite{guo2023pypi}. Registry-side scanning and
continuous-integration tests are defenses. For machine learning (ML), behavioral testing finds
critical failures in natural language processing models~\cite{ribeiro2020checklist}, and simple smoke
tests find severe bugs in ML libraries~\cite{herbold2022smoke}. About 100 curated examples can estimate
full-benchmark accuracy~\cite{polo2024tinybenchmarks}. One conversion study finds that a third of
reported converter failures yield semantically incorrect Open Neural Network Exchange (ONNX)
models~\cite{jajal2024interoperability}. Another finds that converted models keep the originals'
accuracy, although Core ML conversions are more vulnerable to adversarial attacks than ONNX
ones~\cite{openja2022converting}.
In data-quality terms, a model artifact is a data object. The dimension at stake is what Cai and Zhu
call reliability, whether the data can be trusted~\cite{cai2015quality}. Download counts do not establish
it, and small upstream faults compound downstream~\cite{sambasivan2021cascades}.

\section{Study Design}
Fig.~\ref{fig:pipeline} summarizes the pipeline, from a cheap screen over every tag to the rebuild of
confirmed failures from source.

\begin{figure*}[t]
\centering
\begin{tikzpicture}[
  box/.style={draw=black!60, rounded corners=2pt, align=center, font=\footnotesize,
              minimum height=1.25cm, text width=2.35cm, inner sep=3pt},
  arr/.style={-{Stealth[length=2mm]}, black!60}]
\node[box] (a) {\textbf{Census}\\305 official tags\\+ 22 HF files};
\node[box, right=0.42cm of a] (b) {\textbf{15-task screen}\\39 flagged\\(of 305)};
\node[box, right=0.42cm of b] (c) {\textbf{Group baseline}\\7 suspects};
\node[box, right=0.42cm of c] (d) {\textbf{Confirm}\\164 tasks (CUDA);\\Metal: 15 tasks};
\node[box, right=0.42cm of d] (e) {\textbf{Rebuild from source}\\5 suspects; plain\\and calibrated};
\node[box, right=0.42cm of e, text width=2.6cm] (f) {\textbf{Verdict}\\file defect, or\\collapse (rescued\\or not)};
\foreach \u/\v in {a/b,b/c,c/d,d/e,e/f} \draw[arr] (\u) -- (\v);
\end{tikzpicture}
\caption{Study pipeline. Counts: official library, serial scoring. Only the five Qwen
and phi3.5 suspects are rebuilt and replicated on Metal; the two failing community files are tested
separately under generic ChatML and the model's instruction template (RQ2). Unflagged tags are audited on 49 other tasks
(RQ4).}
\label{fig:pipeline}
\end{figure*}

\subsection{Sampling frame}
We enumerated every eligible tag (a named downloadable build) in the official Ollama model families
chosen at design time. A tag is eligible if it is an instruct, code-capable build with an explicit
quantization suffix and a file of at most 8\,GB. This gives 305 artifacts of 35 model variants in 15
model lines, eight code-specialized and seven general instruct (Fig.~\ref{fig:census}). Of these, 291 are
quantized and 14 are unquantized 16-bit floating-point (fp16) builds. Within these lines, this is the
eligible population, not a sample. For the community channel we took 22 files at \texttt{Q2\_K} to
\texttt{Q4\_K\_M} from 12 popular GGUF repositories on HF. The inventory was generated on 10 August 2026
and the census ran on 12 August 2026, with each artifact's pull time logged.

\subsection{A tuned 15-task screen}\label{sec:screen}
We tuned the screen on 24 Qwen2.5-Coder~\cite{hui2024qwen25coder} configurations (0.5B to 7B, 2-bit to
fp16) from our earlier full-benchmark evaluation. Each was scored with EvalPlus~\cite{liu2023evalplus} on
all HumanEval+ and MBPP+ tasks. These extend the tests of HumanEval~\cite{chen2021codex} and Mostly Basic
Python Problems (MBPP)~\cite{austin2021mbpp}. Twenty-three were healthy; the 3B
\texttt{q3\_K\_M} tag, later confirmed non-functional, was the one known failure. Thirty-nine HumanEval+
tasks are solved by all 23 healthy configurations, and the suite is the first 15 of them in lexicographic
order of task identifier. On the tuning set every healthy configuration passes 15 of 15 and the known
failure passes none.

Screening has two stages with fixed constants. Stage one flags an artifact that passes fewer than 9 of 15
tasks. Stage two compares each artifact with its \emph{group baseline}, the median score of the same model
variant (the tag without its quantization suffix, one column of Fig.~\ref{fig:census}) at \texttt{q6\_K},
\texttt{q8\_0} or fp16, and assigns one class:
\begin{itemize}
\item \textbf{healthy}: at least 60\% of the group baseline;
\item \textbf{suspect}: at most 2 of 15 in a group whose baseline is at least 8;
\item \textbf{low baseline}: below 60\% in a group whose baseline is below 8, so the tags score low even at high precision;
\item \textbf{degraded}: any other score below 60\% of the baseline;
\item \textbf{no baseline}: the group has no high-precision tag within the 8\,GB frame.
\end{itemize}
We call a suspect \emph{non-functional} when it also solves under 10\% of the 164 HumanEval+ tasks. Only
suspects proceed to rebuild attribution.

\emph{A scoring fault, corrected.} The census scorer marked HumanEval/15 as failed for every artifact,
although most generations contain the standard solution. Its strengthened tests are slow and can exceed the
time limit during parallel scoring. We therefore re-evaluated every failed task of every result set
serially, one evaluation at a time, with the same EvalPlus version and its default time limits. We also
re-evaluated a 3\% sample of passes, none of which changed. HumanEval/15 then passes for 275 of the 305
official artifacts, one of which also gains HumanEval/23. No task that passed before fails, and in the
192 passing runs we timed, the task's slowest test used at most 20\% of its time limit. The correction
reduces stage-one flags from 50 to 39, leaves all seven suspects unchanged, and moves one non-suspect tag
from degraded to healthy.
We report the serial scores throughout.

\subsection{Attribution by recipe-matched rebuild}\label{sec:attr}
A low score does not identify its cause. Each suspect is rerun on all 164 HumanEval+ tasks on NVIDIA's
CUDA. The five Qwen and phi3.5 suspects are also rerun on the 15-task suite on a second backend (Apple's
Metal GPU interface) and compared with a \emph{plain rebuild}. We download the original checkpoint,
convert it to GGUF and quantize it ourselves at the same level without calibration. This is the recipe
the registry metadata indicates, since it carries no \texttt{quantize.imatrix} keys. We call the
comparison build the suspect's \emph{referee}. We did not rebuild the license-gated llama3.2-1B
checkpoint~\cite{meta2024llama32}, leaving its two non-functional tags unattributed.

The file-defect clause below, the settings checks, the first miss-audit sample and the suite-choice
analysis were fixed in a dated local run plan before any rebuild. After the first rebuild results, we
added the rescue clause in its present form, which compares a calibrated build with our plain rebuild
rather than with the suspect, and entered it in the plan before the controlled-calibration runs. The
rule is:
\begin{itemize}
\item \textbf{file defect} if the suspect scores at most 2 of 15 while our plain rebuild scores at least 5 and at least 4 more;
\item \textbf{collapse} if the plain rebuild fails comparably; the collapse is \textbf{calibration-rescued} when a calibrated build of the same weights at the same level scores at least 5 and at least 4 more than the plain rebuild.
\end{itemize}
Our calibrated rebuilds use our own imatrix and the same checkpoint, llama.cpp commit and type mix as our
plain rebuild. For phi3.5, the file-defect check keeps the tag's mix; calibration is compared at llama.cpp's current mix. We also use
imatrix-calibrated files from the HF uploader bartowski~\cite{hfbartowskiqwen,hfbartowskiphi} and GGUF
files from Qwen~\cite{hfqwengguf} labeled \texttt{Instruct-AWQ} in their metadata. These builds inform
the verdict but never decide a file defect.

We compare each rebuilt suspect with our plain rebuild tensor by tensor, checking type, shape and bytes
while ignoring metadata and tensor order. Different file hashes or repository names do not establish
different conversions. Equality with our rebuild is itself a finding, because we produced it independently
from the original checkpoint. We check the Qwen and phi3.5 third-party referees with tensor hashes and reject any whose
tensors equal the suspect's. The check planned before the runs hashed tensors in file order and initially
reported every plain rebuild as distinct. After the first results, we used the order-independent
comparison reported here.

\subsection{Experimental setup}
Table~\ref{tab:setup} lists the setup and default decoding settings. Default decoding is greedy with a
fixed seed. A repeat on a second RTX 4090 system, using the same Ollama build and task order, reproduced
all 225 answers from 15 census tags byte for byte. Generations are scored with EvalPlus plus-tests,
which require both the base and the strengthened tests to pass. For each generation we also record four
cheap output statistics, namely 6-gram repetition, prompt echo, whether it hit the token cap (the
suite-wide rate is \emph{truncation}), and whether an output line matches the one-line header pattern
\texttt{def name(...):}. The repetition score is six times the count of the most frequent word 6-gram
divided by the word count and can exceed 1. Prompt echo is the fraction of trimmed prompt lines longer than
12 characters that appear verbatim in the output. Community HF files run with no template supplied, so the runtime uses the file's own chat template or its
generic default; other HF files run with the matching official tag's.

\begin{table}[t]
\centering
\caption{Experimental setup.}
\label{tab:setup}
\setlength{\tabcolsep}{3pt}
\footnotesize
\begin{tabular}{@{}lL{6.2cm}@{}}
\toprule
Generation & Ollama 0.32.6~\cite{ollama} (bundled llama.cpp \texttt{96278e3}), pinned for all generation runs \\
Rebuilds & llama.cpp v0.7.0 (\texttt{620d222})~\cite{llamacpp} on CPU: conversion, quantization, imatrix and perplexity \\
Checkpoints & Qwen2.5-Coder-3B/1.5B/0.5B-Instruct~\cite{hui2024qwen25coder} and Phi-3.5-mini-instruct~\cite{abdin2024phi3} at pinned revisions \\
Backends & NVIDIA RTX 4090 with CUDA (census, reruns); Apple M2 with Metal (replication) \\
Decoding & greedy, seed 42, context 2{,}048, at most 768 new tokens \\
Scoring & EvalPlus 0.3.1, HumanEval+ plus-tests, one evaluation at a time \\
Calibration & imatrix from 100 chunks of 512 tokens of WikiText-2 train~\cite{merity2017pointer}, or of 1.5\,MB of Python source from the llama.cpp repository \\
\bottomrule
\end{tabular}
\end{table}

\section{Results}
\subsection{RQ1: how many artifacts are non-functional}
\begin{figure*}[t]
\centering
\includegraphics[width=\textwidth]{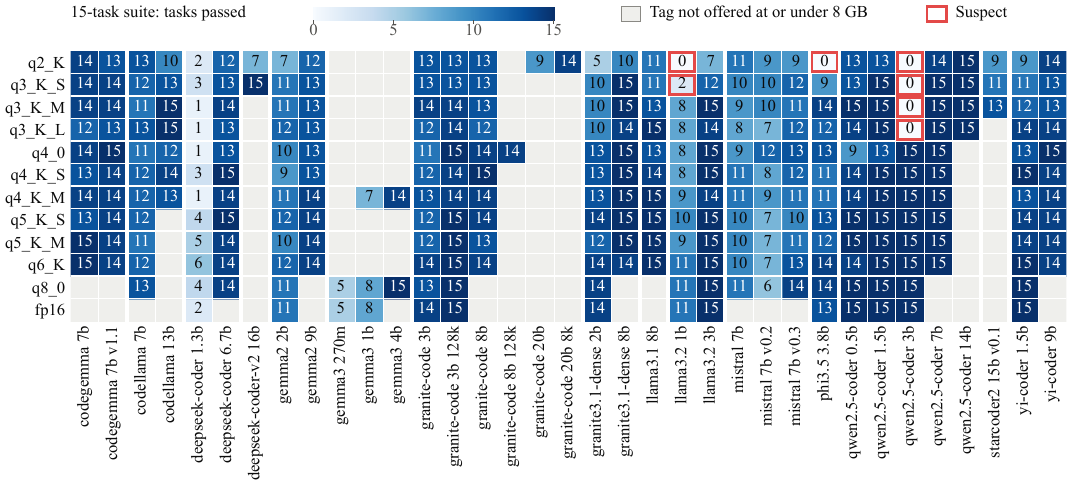}
\caption{15-task suite score (of 15, serial scoring) for all 305 official-library artifacts, one column
per registry model variant and one row per quantization level. Gray: tag not offered at or under 8\,GB.
Red boxes: the seven suspects. The Qwen2.5-Coder-3B column drops to zero at four adjacent low-bit levels
and recovers at 4 bits; deepseek-coder-1.3B scores 1 to 6 at every level, including fp16.}
\label{fig:census}
\end{figure*}

Table~\ref{tab:classes} classifies the 305 official artifacts against their groups' high-precision
baselines, and Fig.~\ref{fig:census} shows every score.

\begin{table}[t]
\centering
\caption{Classification of the 305 official artifacts against per-variant high-precision baselines
(serial scores). All six low-baseline tags, and six of the healthy ones, are deepseek-coder-1.3B
tags scoring 1 to 6 of 15.}
\label{tab:classes}
\begin{tabular}{lr}
\toprule
Class & Count \\
\midrule
Healthy (at least 60\% of group baseline) & 271 \\
Degraded (below 60\%, baseline at least 8) & 2 \\
Low baseline (baseline below 8 of 15) & 6 \\
Suspect (at most 2 of 15, baseline at least 8) & 7 \\
No in-frame baseline & 19 \\
\bottomrule
\end{tabular}
\end{table}

All seven suspects are confirmed on the full benchmark: six solve none of the 164 tasks and the seventh
(llama3.2-1B \texttt{q3\_K\_S}) solves 6. They form three incidents: four Qwen2.5-Coder-3B tags at
adjacent levels, one phi3.5-mini tag and two llama3.2-1B tags. As shares, 7 of the 305 official artifacts
(2.3\%) and 3 of the 35 model variants (8.6\%) are affected. Read as draws from the registry's
conversion process, these give Wilson 95\% intervals of 1.1--4.7\% and 3.0--22.4\%. Tags of one variant
are not independent incidents, so the variant rate is the better summary. All seven sit among the 48 tags
at \texttt{q2\_K} to \texttt{q3\_K\_L} in models of at most 4B parameters (14.6\%, a subgroup we
describe after the fact). None of the other 257 tags is a suspect. In the
community channel, 20 of 22 files are healthy on CUDA; the two exceptions are discussed under RQ2.

One group escapes the relative screen. All 12 deepseek-coder-1.3B tags, fp16 included, score 1 to 6 of
15, and the fp16 tag solves 16 of 164 HumanEval+ tasks, against the 65.2\% on HumanEval that its authors
report under their own prompt~\cite{guo2024deepseekcoder}. Of the group's screen outputs, 59\% do not
parse, about half of them on indentation, against 2.4\% for the 286 official tags outside this group and
the seven suspects. These files lack a pre-tokenizer type, so the runtime tokenizes prompts with a generic
rule (6 to 8\% of prompt tokens differ, mostly indentation and spacing) and warns that generation quality
will be degraded~\cite{gerganov2024bpe}. Yet the deepseek-coder-6.7B and starcoder2-15B tags, the other
14 of the 26 official tags without a pre-tokenizer type, work. We therefore leave the group unattributed.
The count of seven covers failures against a working high-precision sibling and is a lower bound on files that solve at
most 16 of 164: full runs find eight more among non-suspects, seven in this group and gemma3-270M
\texttt{q8\_0}, all of which the proposed gate withholds (Section~\ref{sec:disc}).

The rebuilt failures are stable: the four Qwen files and the phi3.5 file we re-downloaded in October
2026 are byte-identical to the census copies. Independently, an outside reader in the issue
thread~\cite{ollama18252} reproduced the \texttt{q3\_K\_M} failure on a
CPU-only Debian machine with a later Ollama release (0.33.3), recording identical output across runs
and the registry's manifest digest.

\emph{Answer to RQ1:} 2.3\% of the registry's eligible tags are non-functional, all of them low-bit tags
of small models, in three of 35 model variants. A group that scores low at every precision
escapes this count.

\subsection{RQ2: what causes the failures}\label{sec:rq2}
Table~\ref{tab:adjudication} gives the attribution for every suspect.

\begin{table*}[t]
\centering
\caption{Attribution of the seven suspects. Scores are 15-task suite passes (of 15) unless marked /164.
Plain: our uncalibrated rebuild from the original checkpoint; for phi3.5, with the tag's tensor mix and,
in parentheses, with llama.cpp's current mix (21 of 164). Own imatrix: our rebuild with an importance
matrix from WikiText, from code. Third party: bartowski's imatrix builds; for llama3.2-1B, unsloth's files,
in parentheses. Best cal.: best calibrated build on all 164 tasks. Our phi3.5 imatrix builds use the
current mix and score 9 and 8 against 4 for the plain build at that mix; bartowski's keeps the tag's mix
and scores 7 against 0.}
\label{tab:adjudication}
\setlength{\tabcolsep}{4.2pt}
\begin{tabular}{lcccccccl}
\toprule
Library tag & Library & Library /164 & Plain & Own imatrix & Third party & Qwen AWQ & Best cal. /164 & Verdict \\
\midrule
qwen2.5-coder:3b \texttt{q2\_K} & 0 & 0 & 0 & 15, 15 & 15 & 14 & 113 & collapse, rescued by calibration \\
qwen2.5-coder:3b \texttt{q3\_K\_S} & 0 & 0 & 0 & 2, 4 & 4 & -- & 30 & collapse, rescue rule not met \\
qwen2.5-coder:3b \texttt{q3\_K\_M} & 0 & 0 & 0 & 9, 8 & 10 & 15 & 125 & collapse, rescued by calibration \\
qwen2.5-coder:3b \texttt{q3\_K\_L} & 0 & 0 & 0 & 7, 8 & 9 & -- & 86 & collapse, rescued by calibration \\
phi3.5:3.8b \texttt{q2\_K} & 0 & 0 & 0 (4) & 9, 8 & 7 & -- & 37 & collapse, rescued by calibration \\
llama3.2:1b \texttt{q2\_K} & 0 & 0 & -- & -- & (1) & -- & -- & not attributed (no rebuild) \\
llama3.2:1b \texttt{q3\_K\_S} & 2 & 6 & -- & -- & (4) & -- & -- & not attributed (no rebuild) \\
\bottomrule
\end{tabular}
\end{table*}

\begin{figure*}[t]
\centering
\includegraphics[width=0.85\textwidth]{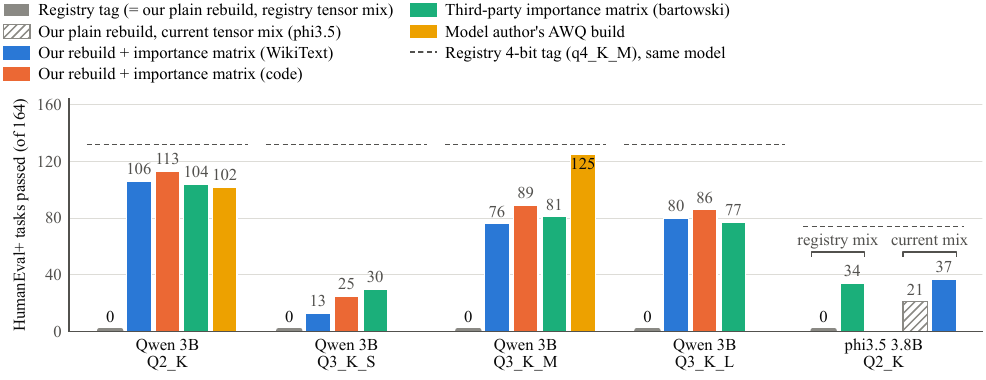}
\caption{Full-benchmark effect of calibration at the levels where the registry tags fail. The gray
marker is the registry tag, which equals our uncalibrated rebuild with the registry's tensor mix (0 of
164 at every level). Colored bars: our calibrated rebuilds and third-party comparisons at the same
level. For phi3.5,
the hatched bar is our uncalibrated rebuild with llama.cpp's current mix (21), the like-for-like
comparator of our WikiText imatrix build (37); bartowski's build keeps the registry's mix (34). Qwen's
AWQ files exist only at \texttt{q2\_K} and \texttt{q3\_K\_M}, no AWQ build exists for phi3.5, and its
code-imatrix build was scored on the 15-task suite only (8 of 15). Dashed: the registry's 4-bit tag of
the same model.}
\label{fig:calibration}
\end{figure*}

\emph{What the failures look like.} The non-functional builds do not crash or refuse; they stream text at
normal speed. Table~\ref{tab:outputs} shows four of their answers to the simplest task in the suite, a
one-line string length function. The registry's \texttt{q3\_K\_M} answer is byte-identical to our plain
rebuild's: fluent preamble, a signature and a docstring, and no body.

\begin{table}[t]
\centering
\caption{Openings of four suspects' answers to HumanEval/23 (\texttt{strlen}), verbatim except that runs
of whitespace are merged; \ldots\ marks omitted text.}
\label{tab:outputs}
\footnotesize
\begin{tabular}{@{}lL{5.6cm}@{}}
\toprule
Tag & Output \\
\midrule
Qwen-3B \texttt{q2\_K} & \texttt{program9 demon1 bil mess(bil is sports Audi program program program\ldots} (to the token cap) \\
Qwen-3B \texttt{q3\_K\_S} & \texttt{\textasciigrave\textasciigrave\textasciigrave python def strlen(s >{}>{}> strlen(\textquotesingle\textquotesingle) \textasciigrave\textasciigrave\textasciigrave{} \textasciigrave\textasciigrave\textasciigrave} \\
Qwen-3B \texttt{q3\_K\_M} & \texttt{. To calculate the length of a given string in Python, . You can use the . Here\textquotesingle s a simple function\ldots{} def strlen(string: str) """Calculate and return the length\ldots{}} \\
phi3.5 \texttt{q2\_K} & \texttt{import \textasciigrave str\textasciigrave("This is a Python script that solves the following problem in various ways.")))} \\
\bottomrule
\end{tabular}
\end{table}

\emph{No file defect.} Rebuilt from the original checkpoints, the four Qwen tags match the library files
in all 434 tensors. Only metadata differs (names, license, sampling defaults and two tokenizer flags), and
library and rebuild give byte-identical answers on all 15 tasks at every level. The phi3.5 tag matches our
rebuild in all 197 tensors once its 32 \texttt{attn\_qkv} tensors are set to \texttt{Q2\_K}, where current
llama.cpp uses \texttt{Q3\_K}, and that rebuild also scores 0 of 15 and 0 of 164. Each of the five files
is therefore an exact, uncalibrated llama.cpp quantization of the released weights, and under our rule
none is a file defect. As a positive control, our plain \texttt{Q4\_K\_M} and \texttt{Q6\_K} Qwen rebuilds
from the same fp16 file score 15 of 15 (\texttt{Q4\_K\_M}: 129 of 164), and our \texttt{Q8\_0} rebuild
matches fp16 perplexity (11.60 against 11.55).

The same runs argue against three simpler explanations. \emph{Tokenizer and template:} our calibrated builds share
the rebuild's tokenizer metadata and chat template, and they work. \emph{Runtime:} the failure appears on
CUDA and Metal under Ollama, in perplexity computed by a separate llama.cpp build on CPU, and on an
outside CPU machine with a later release. \emph{Tied embeddings:} the
working 1.5B model ties its embeddings as the 3B model does, and phi3.5, which does not, also fails.

\emph{Calibration rescues four of five, but only partly.} Fig.~\ref{fig:calibration} shows the
full-benchmark effect of calibration. With weights, toolchain and tensor mix held fixed, our imatrix
takes Qwen \texttt{q2\_K} from 0 to 106--113 of 164, \texttt{q3\_K\_M} to 76--89 and \texttt{q3\_K\_L}
to 80--86. For phi3.5, the current llama.cpp mix lifts our plain build from 0 to 21 of 164 by storing the
32 \texttt{attn\_qkv} tensors at \texttt{Q3\_K} rather than \texttt{Q2\_K} (3.16 instead of 2.96 bits
per weight). The imatrix then lifts it from 21 to 37 (exact McNemar test, $p=0.007$). At
\texttt{Q3\_K\_S} and \texttt{Q3\_K\_M}, which work without calibration, our phi3.5 imatrix builds
also gain, from 35 to 46 and from 61 to 76 ($p=0.07$ and $0.004$; exploratory). bartowski's calibrated
phi3.5 file keeps the tag's mix and solves 34, consistent with rescue at that mix.

For Qwen, bartowski's builds agree at every rescued level (104, 81 and 77), as do Qwen's two files
labeled \texttt{Instruct-AWQ} (102 and 125). The latter also store a separate output head (435 tensors
against 434), so they differ in more than calibration. Qwen \texttt{q3\_K\_S} stays weak under every
calibration (13--30 of 164), and calibrated builds at 7 or 8 of 15 would still be flagged by the screen.
The registry's 4-bit tags score 132 (Qwen) and 74 (phi3.5).

The rescue is weaker at the higher bit widths. Our calibrated \texttt{Q2\_K} builds have about 1.3 times the fp16
perplexity and solve 106--113 tasks, while the calibrated \texttt{Q3\_K\_M} and \texttt{Q3\_K\_L} builds
keep about 6 times the fp16 perplexity and solve fewer tasks. We have not identified the cause of this
remaining gap in the \texttt{Q3\_K} builds. In llama.cpp, supplying an importance matrix also
switches \texttt{Q2\_K} and \texttt{Q3\_K} to a weighted scale search, so our comparison isolates the
imatrix as a whole, not the calibration text alone. The WikiText and code matrices give the same verdicts and
perplexity within 4\%, yet the code matrix solves 6 to 13 more tasks at each Qwen level. Qwen's files,
labeled \texttt{Instruct-AWQ} and carrying no imatrix keys, also work. This is consistent with a
calibration benefit, but their separate output head prevents them from isolating its effect.

\emph{The same recipe does not break 1.5B.} The registry's \texttt{q2\_K} to \texttt{q3\_K\_L} tags of
Qwen2.5-Coder-1.5B also match our uncalibrated rebuild in all 338 tensors and stay functional
(Table~\ref{tab:sizes}: 13--15 of 15), although the \texttt{q2\_K} tag solves only 10 of the 49 RQ4 audit
tasks, against 28 for the \texttt{q4\_0} tag. The 0.5B tags do not test the
recipe. Their 896-wide rows cannot hold 256-weight k-quant blocks, so 96--144 of their 169 weight
matrices fall back to 4-bit types (\texttt{IQ4\_NL} in the library, \texttt{Q4\_0} in our rebuild), and
their \texttt{q2\_K} and \texttt{q3\_K} files run at 5.4--5.9 bits per weight. Bit budget does not explain
the 3B failure either: the registry's uncalibrated 3B \texttt{q4\_0} tag, about the size of the
non-functional \texttt{q3\_K\_L} tag (1.8 against 1.7\,GB), scores 15 of 15. Within this family, then, the
uncalibrated \texttt{Q2\_K} and \texttt{Q3\_K} types stay functional at 1.5B and fail at 3B, and the recipe alone does
not predict the failure. The toolchain does not guard against it: llama.cpp refuses to quantize its very
lowest-bit types without an importance matrix but exempts \texttt{Q2\_K} and the \texttt{Q3\_K} mixes by
design~\cite{llamacppguard}.

\begin{table}[t]
\centering
\caption{The same uncalibrated recipe at two sizes of Qwen2.5-Coder (15-task suite passes): library
tag / our plain rebuild / our WikiText imatrix rebuild. Library and plain tensors are identical at both
sizes.}
\label{tab:sizes}
\setlength{\tabcolsep}{8pt}
\begin{tabular}{lcc}
\toprule
Level & 1.5B & 3B \\
\midrule
\texttt{q2\_K} & 13 / 13 / 12 & 0 / 0 / 15 \\
\texttt{q3\_K\_S} & 15 / 15 / 14 & 0 / 0 / 2 \\
\texttt{q3\_K\_M} & 15 / 15 / 15 & 0 / 0 / 9 \\
\texttt{q3\_K\_L} & 15 / 15 / 14 & 0 / 0 / 7 \\
\bottomrule
\end{tabular}
\end{table}

\emph{Why the referee must match the recipe.} The first version of this study~\cite{patodiya2026v1}
compared the Qwen and phi3.5 suspects with an independent distributor's imatrix-calibrated files
(bartowski), leading to a defective-conversion verdict. Its llama3.2-1B referees also failed, leading to
a collapse verdict. Here we leave those tags unattributed. Distributor identity misleads in another way
too. The phi3.5 library file
is byte-identical to two HF uploads under unrelated names~\cite{quantfactory2024phi35,neopolita2024phi35}
and tensor-identical to a third~\cite{lmkit2024phi35}, so one artifact reaches users through four
storefronts. Such a copy can pass for an
independent referee: file hashes expose the two byte-identical copies, but only a tensor-level
comparison exposes the third.

\emph{The llama3.2-1B pair.} Its only referees, unsloth's files~\cite{hfunsloth}, carry imatrix keys
in their GGUF metadata (file hashes match the revision, tensors not compared) and also fail, at 1
and 4 of 15, so these calibrated builds do not rescue it, and without a rebuild we cannot separate recipe
from model.

\emph{A missing template.} Two deepseek-coder-6.7B files from TheBloke~\cite{hfthebloke}, at
\texttt{Q3\_K\_M} and \texttt{Q4\_K\_M}, score 5 and 6 of 15 on CUDA, while Ollama's own conversions score 14 at
both levels. The files carry no chat template, so the runtime falls back to a generic ChatML format that
differs from the model's documented instruction format: 28 of their 30 answers run into invented chat
turns and 87\% and 96\% of 164-task generations hit the token cap. With the instruction template, the same
files score 13 and 15 of 15 on CUDA and Metal, and their full CUDA scores rise from 16 to 98 and from 18
to 101 of 164. We did not rebuild them and record them as template-dependent screen failures, not file defects.

\emph{Answer to RQ2:} none of the five rebuilt failures is a corrupted file. Four are uncalibrated
quantizations that calibration rescues, at least partly, and Qwen \texttt{q3\_K\_S} stays weak under every
calibration. The llama3.2-1B pair is unattributed, and calibrated third-party builds of it fail too. A
separate community pair recovers with its model's instruction template.

\subsection{RQ3: can failures be detected without executing generated code}
Truncation and repetition catch the noisy failure modes. Truncation at or above 0.2 flags 3 of the 7
suspects and 10 of the other 298 tags. Five of those ten solve at most 16 of 164. Mean 6-gram repetition
at or above 0.2 flags 3 suspects and 4 other tags. Either statistic at 0.3 or above catches 4 suspects
and 6 other tags. Four of those six are deepseek-coder-1.3B tags solving 2 to 16 of 164, a group-wide
failure that escapes the suspect criterion. The rule misses Qwen \texttt{q3\_K\_M},
\texttt{q3\_K\_L} and llama3.2-1B \texttt{q3\_K\_S}. None hits the token cap, and a repetition
threshold low enough to catch any of them also flags at least 4.4\% of healthy tags.

Perplexity does better on the builds we measured (Table~\ref{tab:ppl}), as Singh et al.\ also found for
their unusable Qwen2.5-3B \texttt{q2\_K} build~\cite{singh2026saying}. On WikiText-2, each of the five non-functional library builds we measured has
27 to 5{,}700 times the perplexity of its fp16 model, while the 4-bit tags stay within about 5\% of it (4.2\% and 5.1\%). Thresholds from
10 to 25 times fp16 would flag all five measured non-functional builds without executing generated code,
and our weak calibrated Qwen \texttt{Q3\_K\_S} builds, at 18.6 and 18.8 times, fall inside that range.
We chose the range after seeing these builds. It needs an fp16 reference, and its false-positive rate is
unknown because we did not measure the llama3.2-1B pair or the other tags. Perplexity is not a functional
verdict, though. Our calibrated Qwen \texttt{Q3\_K\_M} build has about 6 times the fp16 perplexity yet
solves 76 of 164 tasks, and perplexity cannot separate a corrupted file from an uncalibrated one.

\begin{table}[t]
\centering
\caption{WikiText-2 test perplexity (20 chunks of 512 tokens; llama.cpp \texttt{620d222} on CPU). Library
low-bit Qwen files equal our plain rebuilds. W / C: importance matrix from WikiText-2 train / Python code;
W is calibrated on the training split of the evaluation corpus; W and C agree within 4\%.}
\label{tab:ppl}
\setlength{\tabcolsep}{4pt}
\begin{tabular}{lrr}
\toprule
Build & Qwen2.5-Coder-3B & phi3.5-mini \\
\midrule
fp16 & 11.5 & 6.9 \\
Library \texttt{q4\_K\_M} & 12.1 & 7.2 \\
Library \texttt{q2\_K} & 66{,}391 & 184.1 \\
Library \texttt{q3\_K\_S}, \texttt{\_M}, \texttt{\_L} & 920, 333, 373 & -- \\
Plain rebuild \texttt{Q2\_K}, current mix & = library & 12.3 \\
Own imatrix \texttt{Q2\_K} (W / C) & 14.4 / 14.9 & 10.4 / 10.5 \\
Own imatrix \texttt{Q3\_K\_S} (W / C) & 214 / 217 & 8.9 / 8.8 \\
Own imatrix \texttt{Q3\_K\_M} (W / C) & 72.7 / 70.8 & 7.7 / 7.7 \\
\bottomrule
\end{tabular}
\end{table}

\emph{Answer to RQ3:} output statistics flag four of seven failures plus six other tags, four of them under
10\% of 164; perplexity flags every
non-functional build we measured, with a threshold chosen after the fact. Neither separates file defects from recipe failure,
which still requires the rebuild comparison.

\subsection{RQ4: does the screen work as an acceptance gate}
Six checks measure the screen's limits.
\begin{itemize}
\item \textbf{Precision.} All seven suspects the screen named solve at most 6 of the 164 tasks.
\item \textbf{Misses.} In the census, only flagged tags received a full evaluation. We therefore sampled from the 233 unflagged official tags outside the tuning set, as scored before the correction: 56 drawn at random in two rounds of up to three from each of ten model-family strata, plus 10 sampled in the first version. Each ran on 49 other HumanEval+ tasks fixed before the audit and solved 8--41, none 2 or fewer. Regardless of its flag, a tag below 10\% solves at most 16 of 164 tasks. Seventeen distinct solved tasks on one file and backend therefore rule out near-total failure without a full run. Fifty artifacts reach 17 on the audit tasks alone, one of them re-published before the audit and tested on its newer file. Fourteen more reach 17 with their screen tasks added, which requires the same file digest, GPU model and Ollama release in both runs. The first-version audit recorded no backend, so we reran the remaining two on the census files. They solve 65 and 44 of all 164 tasks. None of the 66 is non-functional. The audit targets near-total failure, not degradation (e.g., the 1.5B \texttt{q2\_K} tag, Section~\ref{sec:rq2}).
\item \textbf{Population.} The first-version draw is unrecorded, so we bound only the 223 other tags: the 56 seeded draws give an exact stratified 95\% upper bound of 22 non-functional census files (9.9\%), conservatively counting the re-published tag's August file as failed. The release tabulates tighter bounds under further assumptions.
\item \textbf{Suite choice.} On the serial census scores, every subset of at least 3 of the 15 tasks keeps the five rebuilt suspects in the suspect set. The two llama3.2-1B tags are less stable. \texttt{q3\_K\_S}, at exactly 2 of 15, drops out of 31--87\% of subsets at every size from 3 to 14. Most dropouts occur because the suspect cap, scaled to $2k/15$ for a $k$-task subset, falls below its subset score. \texttt{q2\_K} drops out of at most 22\%, only at 8 or fewer tasks. The stage-one flag set overlaps the full suite's with a mean Jaccard index of 0.63 at 5 tasks and 0.78 at 10. Drawing 15 tasks from all 164 would flag at least one healthy tuning configuration in 94\% of suites, hence our use of the always-solved pool.
\item \textbf{Generation settings.} Every verdict we re-tested holds under the settings in Table~\ref{tab:settings}; calibrated builds were not rerun under them.
\item \textbf{Cost.} Screening one artifact takes 12 to 105\,s of generation on one consumer GPU (median 30\,s), excluding download.
\end{itemize}

\begin{table}[t]
\centering
\caption{15-task suite passes (of 15) under alternative generation settings. Base: greedy, context
2{,}048, 768-token cap. T0.2: temperature 0.2 with seeds 1, 2 and 3. Lib.: registry library tag; suspects: the five rebuilt ones. Plain Qwen-3B: our \texttt{Q2\_K} and \texttt{Q3\_K\_M} rebuilds.}
\label{tab:settings}
\setlength{\tabcolsep}{2.4pt}
\footnotesize
\begin{tabular}{@{}lccccc@{}}
\toprule
Build & Base & Ctx 4k & Ctx 8k & Cap 2k & T0.2 \\
\midrule
Lib.\ suspects, plain Qwen-3B & 0 & 0 & 0 & 0 & 0, 0, 0 \\
Plain phi3.5 \texttt{Q2\_K} (current mix) & 4 & 4 & 0 & 2 & 3, 4, 3 \\
Lib.\ Qwen-3B \texttt{q4\_K\_M} & 15 & 15 & 15 & 15 & 15, 15, 15 \\
Lib.\ phi3.5 \texttt{q4\_K\_M} & 11 & 11 & 11 & 9 & 11, 11, 11 \\
\bottomrule
\end{tabular}
\end{table}

\emph{Answer to RQ4:} all seven suspects the screen named solve at most 6 of the 164 tasks. For each of
the 66 audited tags, a tested file solves at least 17, so none is non-functional. Screening takes about 30\,s of
generation per tag; separating file defects from recipe failure also needs the rebuild.

\section{Discussion}\label{sec:disc}
\textbf{A registry acceptance gate.} We propose the following publication procedure based on these
results:
\begin{itemize}
\item \textbf{Pre-check without executing generated code} where an fp16 build exists: flag a tag whose WikiText-2 perplexity is many times that of the fp16 build (Table~\ref{tab:ppl}); the threshold still needs tuning on more models.
\item \textbf{Screen} the tag on the 15-task suite (about 30\,s of generation on one consumer GPU) and publish it if it solves at least 9 and the pre-check does not flag it (Section~\ref{sec:screen}).
\item \textbf{Confirm} any flagged tag that is not a suspect using the 164-task benchmark. Publish it only if it solves 17 distinct tasks. The run can stop once it reaches 17. Of the 16 such census tags with full runs, the gate would publish 8 and withhold 8, including seven deepseek-coder-1.3B tags.
\item \textbf{Attribute} a suspect by rebuilding it without calibration from the source checkpoint and comparing tensors. Build a calibrated version at the same level and put it through the gate. If it fails the gate, withhold the level. Record the recipe and any calibration input in the file metadata.
\end{itemize}
The quantcheck release implements the screen and scorer and includes our rebuild, calibration and
tensor-comparison scripts.

\textbf{For registries.} The registry served five non-functional files unchanged, byte for byte, from
our August 2026 census to October 2026. Its library pages gave users no failure warning and stated
neither the quantization recipe nor whether a tag was tested. The first step is cheap: screen every
low-bit tag at publication. A suspect needs the diagnosis above, after which the registry can publish a
calibrated build at that level or withhold it. Each model and recipe pair needs its own test. The same
uncalibrated recipe works at 1.5B and fails at 3B, while nominally low-bit 0.5B files fall back to 4-bit
types. Tag metadata should record effective bits per weight as well as the level name. Recording the
recipe and any calibration input would also make tags verifiable~\cite{liu2026verifiability}. Qwen's
own Qwen2.5-Coder-3B GGUF files carry \texttt{Instruct-AWQ} labels and include only two of the four
low-bit levels~\cite{hfqwengguf}, consistent with its guide~\cite{qwen2025guide}. The registry's tags
for this model are uncalibrated and include all four.

\textbf{For practitioners.} Test a downloaded artifact before relying on it, especially a low-bit tag of a
small model, and test it with the backend and chat template you will deploy with. The 15-task suite flags every failure
reported here; a perplexity check caught the five we measured.

\textbf{For research.} Studies that use registry tags as stand-ins for a model at a given precision can
misread a recipe problem as a precision limit. SynConfRoute measured 0\% for Qwen2.5-Coder-3B at
\texttt{q2\_K} via Ollama on single-line infilling and recommends \texttt{Q4\_K\_M} for models of 3B or
more~\cite{thangarajah2026synconfroute}. Our calibrated \texttt{Q2\_K} builds solve about two thirds of
HumanEval+, although we have not confirmed which file SynConfRoute used.

\section{Threats to Validity}
\begin{itemize}
\item \textbf{Construct.} The screen tests Python code generation only, so counts describe that use. An artifact can fail other tasks and still pass. Every healthy Qwen2.5-Coder tuning configuration solves the suite's tasks. They are therefore easy by construction and were selected using the model family where four of the seven failures occur. On MBPP+, the 23 healthy tuning configurations solve 39--70\% of 378 tasks, while the known failure solves 7 tasks. The 15-task threshold is coarse. For phi3.5, calibration raises the score from 4 to 9 of 15 at the same tensor mix. Its 164-task results support the same calibration-rescue verdict (Fig.~\ref{fig:calibration}).
\item \textbf{Internal.} Verdicts could reflect errors in the evaluation harness. Healthy siblings run in the same pipeline, and the five rebuilt suspects score 0 of 15 on both CUDA and Metal, with different outputs on each. The llama3.2-1B pair was neither rebuilt nor replicated on Metal. The HumanEval/15 scoring fault shows that harness errors happen; correcting it changed no suspect or attribution verdict. The attribution rules were recorded in a dated local run plan, included in the release. The plan postdates the public calibration hypothesis~\cite{ollama18252} and was not publicly preregistered. Section~\ref{sec:attr} distinguishes the file-defect rule specified before rebuilding from the rescue rule added after the initial results.
\item \textbf{External.} The census covers one registry's code-capable instruct tags at or under 8\,GB in 15 model lines and a small, popularity-selected community sample. The count of seven is a lower bound because groups that score low at every precision, such as deepseek-coder-1.3B, are not counted. The rescue evidence comes from Qwen2.5-Coder-3B and phi3.5, with most results from Qwen. Whether other models fail under uncalibrated low-bit recipes remains open. Our generation runs used one pinned runtime, and our rebuilds use a newer llama.cpp than the registry probably did.
\end{itemize}

\section{Conclusion}
A quantized model file can be reproduced exactly from its source weights and still not work. In the
official Ollama library, seven of 305 eligible tags are non-functional. Our uncalibrated rebuilds
matched five of these files tensor for tensor, so none of them is corrupted. For Qwen2.5-Coder-3B, a
single calibration step lifts three of the four failing levels from zero to 46--69\% of HumanEval+.
The same uncalibrated recipe leaves its 1.5B sibling working, showing that the recipe alone does not
predict failure.

These results separate two properties that are easy to treat as one. Provenance and reproducibility
checks show where a file came from and how it was made. Only functional testing shows whether it
performs the intended task. Our 15-task screen needs about 30\,s of generation per tag, and none of the
66 sampled tags it passed is non-functional. Yet the registries we examined document no such test. We
recommend that registries test every low-bit tag before publication and report the result alongside
its quantization recipe. Users can then see both how the file was made and how it performed.

\section*{Acknowledgment}
We thank GitHub user \texttt{mitoyosh} for reproducing a failure on hardware we did not have and for
raising the calibration hypothesis tested here. AI assistance (Claude, Anthropic) was used to draft and
edit text in all sections, to write the analysis, figure and experiment scripts, and to run the
experiments. The author designed the study, reviewed all text and results, and takes full responsibility
for the content.

\section*{Data Availability}
The audit dataset (transcripts, output statistics, verdicts and scores), the rebuild and referee file
hashes, per-tensor comparisons and run logs, the dated run plan, and the quantcheck tool are available at
\url{https://github.com/aditi-p31/quantcheck}, release \texttt{v3.4}, together with the per-task results of
the 24 tuning configurations, from which the scripts recompute the 39-task pool, the suite, the MBPP+
figures and the 94\% statistic. The README describes the unreleased inputs, among them the census server
log and the tuning study's original evaluation files, and how the scripts handle their absence.

\bibliographystyle{IEEEtran}
\bibliography{references}

\end{document}